\documentclass[usenatbib]{mn2e}
\usepackage{graphicx}
\usepackage{pslatex}
\usepackage{multirow}
\usepackage{float}
\usepackage{wasysym}
\def\deg{{\rm o}}

\author[B. Gauza et al.]{
B. Gauza$^{1,2}$~\thanks{e-mail:bgauza@iac.es},
V.~J.~S. B\'{e}jar$^{1,2}$,
R. Rebolo$^{1,2,3}$, 
K. Pe\~{n}a Ram\'{i}rez$^{1,2}$,
M.~R. Zapatero Osorio$^4$,
\newauthor
\hspace*{1.2mm}A. P\'{e}rez-Garrido$^5$,
N. Lodieu$^{1,2}$,
D.~J. Pinfield$^6$,
R.~G. McMahon$^{7,8}$,
E. Gonz\'{a}lez-Solares$^7$, 
\newauthor
\hspace*{1.2mm}J.~P. Emerson$^9$,
S. Boudreault$^{1,2}$,
M. Banerji$^7$
\\
\\
$^1$Instituto de Astrof\'{i}sica de Canarias (IAC), E-38200 La Laguna, Tenerife, Spain\\
$^2$Dept. Astrof\'{i}sica, Universidad de La Laguna (ULL), E-38206 La Laguna, Tenerife, Spain\\
$^3$Consejo Superior de Investigaciones Cient\'{i}ficas, Madrid, Spain \\
$^4$Centro de Astrobiolog\'{i}a (CSIC-INTA), E-28850 Torrej\'{o}n de Ardoz, Madrid, Spain\\
$^5$Universidad Polit\'{e}cnica de Cartagena, Campus Muralla del Mar, Cartagena, Murcia E-30202, Spain\\
$^6$Centre for Astrophysics Research, Science and Technology Research Institute, University of Hertfordshire, Hatfield AL10 9AB\\
$^7$Institute of Astronomy, University of Cambridge, Madingley Road, Cambridge CB3 0HA, UK\\
$^8$Kavli Institute for Cosmology, University of Cambridge, Madingley Road, Cambridge CB3 0HA, UK\\
$^9$Astronomy Unit, School of Physics \& Astronomy, Queen Mary University of London, London, E1 4NS, UK.
}
\begin{document}
%
\title[A new L-dwarf member of the HD\,221356 system]{A new L-dwarf member of the moderately metal-poor triple system HD\,221356}
%
\maketitle

\begin{abstract}
We report on the discovery of a fourth component in the HD\,221356 star system, 
previously known to be formed by an F8V, slightly metal-poor primary ($\left[
\rm{Fe/H}\right]=-0.26$), and a distant M8V\,+\,L3V pair. In our ongoing 
common proper motion search based on VISTA Hemisphere Survey (VHS) and 2MASS 
catalogues, we have detected a faint $J=13.76\pm0.04$\,mag) co-moving 
companion of the F8 star located at angular separation of $12.13\pm0.18$\,arcsec 
(position angle of 221.8$\pm1.7$\,deg), 
corresponding to a projected distance of $\sim$312\,au at 26\,pc. Near-infrared
spectroscopy of the new companion, covering the 1.5--2.4 micron wavelength range 
with a resolving power of R$\sim$600, indicates an L1$\pm$1 spectral type. Using 
evolutionary models the mass of the new companion is estimated at $\sim$0.08 
solar masses, which places the object close to the stellar-substellar borderline. 
This multiple system provides an interesting example of objects with masses 
slightly above and below the hydrogen burning mass limit. The low mass 
companions of HD\,221356 have slightly bluer colours than field dwarfs 
with similar spectral type, which is likely a consequence of the sub-solar 
metallicity of the system. 
\end{abstract}

\begin{keywords}
   stars---low-mass, brown dwarfs---stars---individual: HD 221356
\end{keywords}
%
\section{Introduction}
%
Because of progressive cooling with age, brown dwarfs do not obey a unique 
mass-luminosity relation \citep{1997ApJ...491..856B, 2001RvMP...73..719B}.
Therefore, the determination of a brown dwarf mass requires either a good 
knowledge of its age or a direct dynamical measurement. This, in turn, is 
possible for substellar companions of stars or for those that are found in 
multiple systems. An additional advantage is that the metallicity can be 
inferred from the primary star. For solar-type stars the atmospheres are 
much better understood than for very low-mass stars and brown dwarfs, given 
the poor knowledge of opacities in cool atmospheres \citep{2006ApJ...652.1604B, 
2005A&A...442..635B}. Coeval systems containing low-mass companions also 
provide very useful constraints on evolutionary and atmospheric models 
(e.g. \citealp{2006MNRAS.368.1281P, 2010ApJ...721.1725D}) as well as 
offering a rather unique view on how the process of star formation works 
at the very bottom of the main sequence \citep{2007prpl.conf..427B}. In 
particular, brown dwarf companions with well determined metallicities are 
benchmark objects allowing for a better understanding of the effects of 
metallicity on the physical properties and evolution of substellar objects 
\citep{2012MNRAS.422.1922P}. Unfortunately, substellar companions located 
at wide separations ($>$50\,au) from their parent stars are relatively 
rare, with an estimated frequency of less than a few per cent 
\citep{2004AJ....127.2871M, 2007ApJ...670.1367L, 2007ApJ...662..413K}. 
 
We are conducting a search for very low-mass common proper motion companions 
of nearby ($\apprle 25$\,pc) stars, using the VISTA Hemisphere Survey (VHS) 
(McMahon et al., in preparation) and Two Micron All Sky Survey (2MASS) 
\citep{2006AJ....131.1163S}. Our sample of objects includes some of the known 
multiple systems (\citealp{2010AJ....139..176F} and references therein). One 
of the targets investigated so far was HD\,221356, already known to be a 
triple system. The F8.0 V primary is a field star with slightly subsolar 
metallicity $\left[\rm{Fe/H}\right]=-0.26$ \citep{2005ApJS..159..141V}, 
located at $26.12\pm0.37$\,pc \citep{2007AA...474..653V}. The main properties 
of this star are given in Table~\ref{tab:tab1}. \cite{2000AJ....120.1085G} 
reported that the secondary, initially described as a single object, has a 
photometric distance consistent with that of the HD\,221356 star determined 
by {\it Hipparcos}. The secondary was later resolved by \cite{2002ApJ...567L..53C} 
into a binary separated by $0.57$\,arcsec ($\sim$14.9 au), using adaptive 
optics on the Gemini North Telescope. Based on their photometric colours, 
they estimated spectral types of M8.0\,V and L3.0\,V for each of the two 
components, respectively. This binary (hereafter referred to as HD\,221356BC) 
was also investigated by \cite{2007ApJ...667..520C}. Using data at epochs 
separated by 48.3 years, he confirmed a common proper motion between
HD\,221356A and BC. He also measured a mean separation of 
$\rho=451 \farcs 8 \pm 0 \farcs 4$ between both components, which corresponds 
to a projected physical separation of nearly twelve thousand au, making it 
one of the widest known systems with an L-type component (see also fig 11 of 
\citealp{2010MNRAS.404.1817Z}).

In this article we present the identification and characterization of a fourth, 
very low-mass companion of the HD\,221356 system. We outline the procedure and 
results of our proper motion search together with the analysis of $I$ and 
$YJHK_s$-band photometry and near-infrared spectroscopic data of the identified 
companion. We derive the physical properties of the new object which turns out 
to be very close to the hydrogen burning mass limit.   

\begin{table}
\caption{Properties of HD\,221356A (a.k.a. HR\,8931, HIP\,116106).}
\centering
\begin{tabular}{l l c}
\hline
\hline
Parameter & Value & References\\
\hline
RA. (J2000) &  $23^{\rm h} 31^{\rm m} 31 \fs 62$ & -\\
Dec. (J2000) &  $-04\degr 05\arcmin 16 \farcs 78$ & -\\
$V$ (mag) & 6.50 & 1\\ 
Spectral type & F8.0 V & 3 \\
$\mu_{\alpha}\cos(\delta)$ & $178.7 \pm 0.9 $ mas/yr & 2\\
$\mu_{\delta}$ & $-192.8 \pm 0.9$ mas/yr & 2\\
Parallax & 38.29 $\pm$ 0.54 mas & 2\\
Distance & 26.12 $\pm$ 0.37 pc & 2\\
T$\rm{_{eff}}$ & 5976 $\pm$ 44 K & 1 \\
$\log(g)$ & 4.31 $\pm$ 0.06 cm/s$^2$ & 1 \\
Mass & 0.94 $\pm$ 0.13 M$_{\odot}$ & 1 \\
$\left[\rm{Fe/H}\right]$ & $-0.26 \pm 0.03$ & 1 \\
Age & $2.5-7.9$ Gyr & 1\\
\hline
1) \cite{2005ApJS..159..141V} & 2) \cite{2007AA...474..653V} \\
3) \cite{2007ApJ...667..520C}  \\
\hline
\end{tabular}
\label{tab:tab1}
\end{table}

%
\section{Identification and follow-up observations}
%
\subsection{VISTA Hemisphere Survey data}
The new low-mass companion of HD\,221356\,A was identified using the 2MASS and VHS 
catalogues. The VHS is a near-infrared public survey 
intended to cover the entire Southern hemisphere ($\sim20,000~\rm deg^2$) in 
the $JK_s$ broad band filters with a sensitivity more than 3 mag deeper than 
2MASS. It uses the 4.1-m telescope VISTA (Visible and Infrared Survey Telescope 
for Astronomy) operating since 2009 at ESO's Cerro Paranal Observatory in Chile 
\citep{2006Msngr.126...41E}. It is equipped with a wide-field near-infrared 
camera (VIRCAM), comprising sixteen 2k$\times$2k pixel detectors with a mean 
plate scale of $0.339$\,arcsec. The HD\,221356 system was observed with VISTA on 
2010 November 25 and 26. Average seeing conditions were $1.4^{\prime\prime}$ 
and $0.9^{\prime\prime}$, respectively.

The VHS near-infrared images are processed and calibrated automatically by a 
dedicated science pipeline implemented by the Cambridge Astronomical Survey Unit (CASU). 
Standard reduction and processing steps include dark and flat-field corrections, 
sky background subtraction, linearity correction, destripe and jitter stacking. 
For more detailed description we refer to the CASU webpage http://casu.ast.cam.ac.uk/surveys-projects/vista, 
as well as to \cite{2004SPIE.5493..411I} and \cite{2010ASPC..434...91L}.  

\subsection{Proper motion}

The search for additional common proper motion companions of HD 221356 A 
was done using the astrometry given in the VHS and 2MASS catalogues, which provide 
a 12.18 yr baseline. The positions of 2MASS sources have an estimated accuracy 
of 70-80 mas over the magnitude range of $9<K_s\leq16$ \citep{2006AJ....131.1163S}. 
The astrometric solution for VHS observations is done automatically as part of the 
CASU pipeline, using the 2MASS point source catalogue. The objects on the catalogues 
extracted from each VISTA detector are matched to their counterparts in 2MASS using a 
correlation radius of 1 arcsec. Because 2MASS has a high degree of internal 
consistency it is possible to calibrate the world coordinate system of VISTA 
images to better than 0.1 arcsecond.

\begin{figure}
\centerline{
\vbox{
\hbox{
\includegraphics[keepaspectratio=true,scale=0.75]{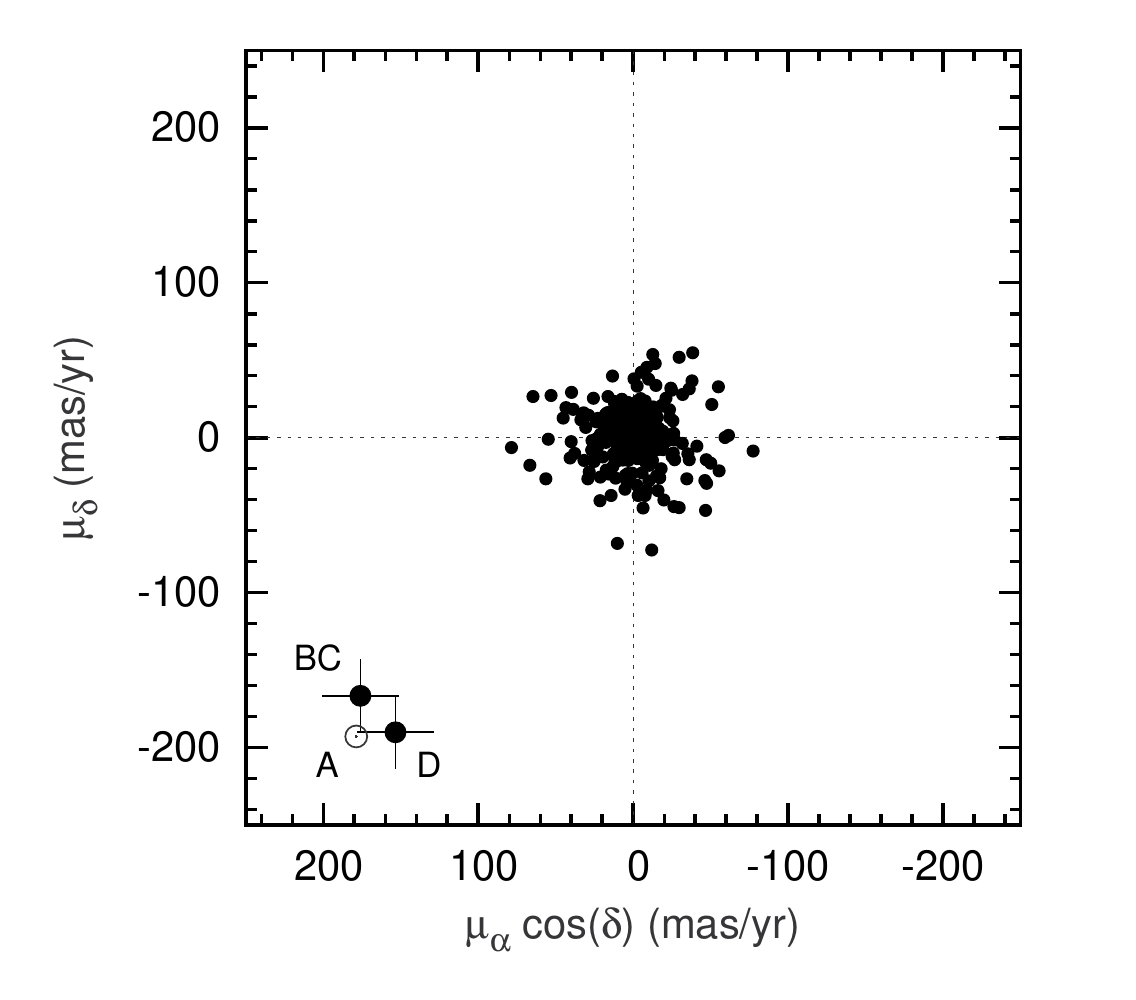}
}}}
\caption{Proper motion vector-point diagram for the HD\,221356 system. All correlated 
objects within 15 arcmin from the primary are plotted as black dots, with HD\,221356 
components labelled as A, BC and D. The primary is saturated in both surveys, its 
proper motion value was taken from the literature. Time baseline between the 2MASS 
and VHS epochs is 12.18 years.}
\label{fig:1}
\end{figure}

The search was performed using TOPCAT,\footnote[1]
{http://www.star.bris.ac.uk/$\sim$mbt/topcat/} a useful tool for analysis 
and manipulation of source catalogues and other data tables, developed as part of the Virtual Observatory. 
We retrieved the astrometric and photomeric data from both 
2MASS and VHS catalogues, for all the objects within a radius of 15 arcmin corresponding 
to $\sim$23,000 au around the examined star. To avoid some of the spurious 
detections, we selected sources brighter than $J$=17 mag in 2MASS.
We have cross-matched 300 objects from both catalogues within 
1 arcsec. The sources that remained unmatched were subsequently cross 
correlated taking into account the proper motion of the primary star 
provided by {\it Hipparcos} \citep{2007AA...474..653V}. We illustrate 
the resulting proper motion vector-point diagram of HD 221356 on Figure \ref{fig:1}.

We have found that the proper motion of HD 221356 BC (Table~\ref{tab:tab2}) is
consistent with that of the primary HD 221356 A, thereby confirming the result of the 
K\"{o}nigstuhl survey of \cite{2007ApJ...667..520C}. Individual 
components of the BC pair, separated by only 0.57\,arcsec \citep{2002ApJ...567L..53C}, 
were not resolved in the VHS images. We also identified a new common 
proper motion companion (2MASS J23313095-0405234, hereafter HD\,221356\,D), with $(\mu_{\alpha}\cos\delta,~\mu_{\delta})
=(153.48\pm21,-190.20\pm19)$ mas/yr, located $12.13\pm0.18$\,arcsec southwest from 
the primary. We adopted a total astrometric uncertainty of $\sim 28$ mas/yr, 
estimated using the standard deviation of proper motions for sources with $\mu\,<$100 mas/yr. 
The proper motion of the new object is common with that of the HD 221356 system. 
Measured separations, position angles and proper motions are listed in Table~\ref{tab:tab2}.

\subsection{Photometry}

The VHS catalogue provides aperture photometry for HD\,221356 in the $YJHK_s$ near-infrared 
bands. The VISTA photometric system is calibrated using the magnitudes of colour selected 2MASS 
stars converted onto the VISTA system using colour equations, including terms to account 
for interstellar reddening\footnote[2]{http://casu.ast.cam.ac.uk/surveys-projects/vista/technical/photometric-properties}. 
Photometric calibrations are determined to an accuracy of 1-2\%. 

 \begin{figure}
 \centerline{
\vbox{
\hbox{
 \hbox{\includegraphics[scale=1]{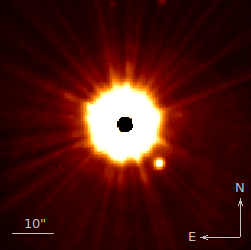} }
}}}
\caption{False colour VISTA $J$-band image of HD 221356AD. Angular separation 
is $12.13\pm0.18$ arcsec and the position angle of the identified companion  
 is $221.8\pm1.7$ degrees. Saturation in the centre of the primary is visible. Field-of-view is 
1$\times$1 arcmin, with North up and East to the left.}
\label{fig:2}
\end{figure}

The new faint companion HD 221356 D is well resolved in the VHS images (see Figure
~\ref{fig:2}), but within the glare of the primary. In order to minimize the 
possible light contamination in the aperture photometry of the VHS catalogue, we  
applied a method to suppress the PSF of the primary by subtracting the flipped 
and rotated images from the original ones. Standard deviation of the background 
at the same separation as the companion was a factor two lower in the PSF 
subtracted images than in the original ones, which slightly improves the 
detection of the object. We performed aperture photometry 
on the resultant images with an aperture of one full-width half-maximum (FWHM). 
The instrumental magnitudes were then calibrated to apparent magnitudes by adding 
the aperture corrections determined using the VHS photometry of fourteen isolated 
bright stars located within 10 arcmin from HD 221356A. The differences between 
the new and the catalogue photometry in the $Y, J, H$ and $K_s$ bands are 
$0.53\pm0.06$, $0.26\pm0.04$, $0.15\pm0.03$, $0.13\pm0.03$ mag, respectively. 
Photometric values for the BC companion were taken directly from the VHS source catalogue. 

\begin{table} \scriptsize
\caption{Proper motion, separations and position angles of low mass components of the HD 221356 system.}
\centering
\begin{tabular}{c c c c c c c}
\hline
\hline
\multirow{2}{*}{Comp.} &  $\mu_{\alpha}\cos(\delta)$  & $\mu_{\delta}$ & $\rho$*   & $\theta$* & $r$  \\
                      &  (mas/yr)                    & (mas/yr)       & (arcsec) & (deg)    & (au) \\
\hline
 BC & $176\pm21$&$-167\pm19$ & $451.10\pm0.18$ & $261.77\pm0.04$ & $11900\pm50$ \\
 D & $154\pm21$&$-190\pm19$ & \hspace*{0.7mm}$12.13\pm0.18$ & $221.8\pm1.7$ & \hspace*{1.2mm}$317\pm9$  \\
\hline
\end{tabular}

\raggedright \hspace*{2mm} * epoch (MJD) = 55525.12460836; $\rho$, $\theta$ and $r$ are measured with respect to the primary.
\label{tab:tab2}
\end{table}
 
On November 15th, 2011 we performed follow-up $I$-band imaging of HD 221356 AD. 
Observations were carried out using the IAC80 telescope equipped with a E2V 
2048$\times$2048 back illuminated CCD detector with a plate scale of 0.304 
arcsec/pix, which provides a 10.4$\times$10.4 arcmin field of view. 
We selected the 12 images with best seeing (FWHM$<$1.2 arcsec) to minimize interference from the primary star.
We reduced the images applying standard techniques, including bias and 
flat-field correction, using \textsc{iraf} routines. Individual exposure times were 5s. For each image, we  
performed a similar method for the PSF subtraction of the primary as used 
for VISTA images and subsequently aligned and combined all of them. 
We obtained the PSF-fit photometry using the \textsc{daophot} package in \textsc{iraf} and calibrated the instrumental 
magnitude of our object using 11 bright stars in the field with DENIS \citep{1999A&A...349..236E} $I$-band data available.  
We note that the photometric system used (Cousin) is not the same as the one of DENIS, 
and that some differences may appear for very cool objects, however in our previous photometric 
calibrations we found that the zeropoint between IAC80 and Denis $I$-band has small colour dependence \citep{2005A&A...443.1021C}.

Additionally, we acquired $I$-band observations of HD\,221356\,A using FastCam, 
mounted on the 1.5-m Carlos S\'{a}nchez Telescope at the Teide Observatory on January 31, 2012. 
FastCam is a lucky imaging instrument, designed to perform high spatial 
and time resolution observations \citep{2008SPIE.7014E.137O}. Optics provide 
a plate scale of 43.5 mas/pix, and a field of view of $\sim22\times22$ 
arcsec$^2$. We obtained 17 blocks of 1000 images of 50\,ms individual exposure 
times. Images were bias corrected, aligned and stacked into the final image 
using the software provided by the FastCam team. We derived 
the $I$-band aperture photometry of the primary, since we found that the 
literature values based on photographic plates are not reliable. Instrumental 
magnitudes were calibrated using photometric standard stars from \cite{1992AJ....104..340L} 
observed at different airmasses along the night under photometric conditions.  
We also explored the inner region to search for the presence of
additional companions, but none were identified. We may exclude the presence 
of an equal mass companion to the primary at separations greater than 
0.2$^{\prime\prime}$ ($\sim$5\,au) and companions with $\Delta I < 5$\,mag at separations 
greater than 1$^{\prime\prime}$ ($\sim$26\,au).
   
The photometric data are listed in Table~\ref{tab:tab3}. The $I$-band 
magnitudes of the distant BC pair were taken from \cite{2003AJ....125.3302G} 
who imaged individual components with Hubble Space Telescope WFPC2. The 
integrated $JHK_s$ photometry of the BC component from VHS were 
also decomposed into individual magnitudes using the flux ratios derived by 
\cite{2002ApJ...567L..53C}. The $JHK_s$ band photometry of the primary, 
also given in Table~\ref{tab:tab3}, is from the 2MASS catalogue. 

\begin{table*}
\caption{Photometric data for the components of the multiple system HD 221356.}
\centering
\begin{tabular}{c c c c c c}
\hline
\hline
Component & $I$ (mag) & $Y$ (mag) & $J$ (mag) & $H$ (mag) & $K\rm{_s}$ (mag) \\
\hline
 A &  5.953$\pm$0.018 & ... & 5.488$\pm$0.019 & 5.264$\pm$0.038 & 5.150$\pm$0.017  \\
BC & 15.536$\pm$0.027 & 13.695$\pm$0.025 & 12.852$\pm$0.010 &12.261$\pm$0.026 &11.946$\pm$0.026  \\
 B & 15.568$\pm$0.018 & ... & 12.933$\pm$0.011 & 12.353$\pm$0.026 & 12.055$\pm$0.028  \\
 C & 19.363$\pm$0.039 & ... & 15.713$\pm$0.042 & 14.993$\pm$0.061 & 14.495$\pm$0.114  \\
 D & 16.70$\pm$0.10 & 14.933$\pm$0.059 & 13.763$\pm$0.038 & 13.209$\pm$0.026 & 12.755$\pm$0.025 \\
\hline
\multicolumn{6}{p{.665\textwidth}}{\footnotesize Notes: $I$-band mag of A is from the FastCam, $JHK_s$ mag are from 2MASS. 
$I$ mag of the BC is from \protect\cite{2003AJ....125.3302G}, $JHK_s$ mags are from VHS and were decomposed using the flux 
ratios derived by \protect\cite{2002ApJ...567L..53C}. $I$ mag of D is from IAC80 measurement, $JHK_s$ mags are from our photometry on VHS images.}
\end{tabular}
\label{tab:tab3}
\end{table*}

\begin{figure*}
\centering
\includegraphics[scale=0.77]{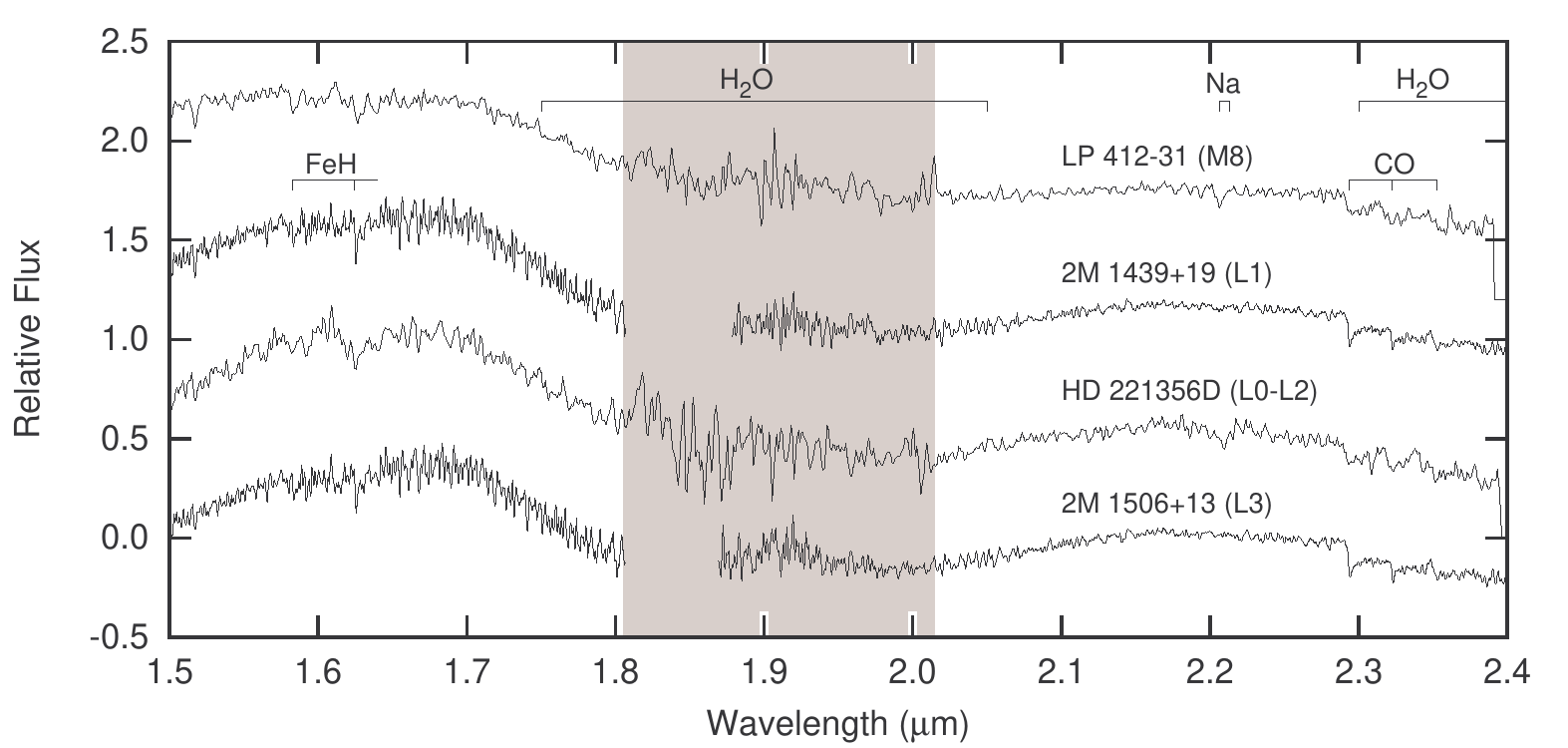}
\caption{Near-infrared $H$- and $K_s$- band spectra of the new companion compared 
with the M8 standard LP 412-31 \citep{1995AJ....109..797K}, observed with the same instrumentation, and the L1 (2MASS\,J14392836+1929149) 
and L3 templates (2MASS\,J15065441+1321060) taken from the IRTF library \citep{2005ApJ...623.1115C}.  
Spectra were normalised at 1.7 microns and offsets have been added for clarity. The grey area
indicates the region of high telluric absorption. The most prominent molecular and 
atomic features are indicated.}
\label{fig:3}
\end{figure*}

\subsection{Near-IR spectroscopy}
We also obtained near-infrared spectroscopy of the HD\,221356 AD system and 
the M8V spectroscopic standard LP 412-31 \citep{1995AJ....109..797K} using LIRIS, with the 
$HK$ grism and the 1K$\times$1K Hawaii detector at the 4.2 m William Herschel 
Telescope (WHT) on December 30 2011. This instrumental configuration provides 
a nominal dispersion of 9.7\,\AA\,pix$^{-1}$ and a wavelength coverage of 
1.4--2.4~$\mu$m. A slit width of 0.75" was used rotated to the direction 
along the AD system and the final resolution of the spectrum was 26\,\AA (R\,$\sim$\,600). 
Total integration time was 2240\,s, divided into individual 
exposures of 160s. A nodding pattern of two positions (AB) was used to subtract 
the sky background. Weather conditions were photometric and the average seeing 
was 0.9". Data were dark corrected, sky subtracted, aligned and combined at 
each nodding position.
Flat-field correction using a tungsten lamp was not applied due to a spurious feature in the $K$-band tungsten spectrum.
After subtracting the contribution of the primary wings, spectra of HD\,221356\,D 
were optimally extracted using the \textsc{apall} routine and wavelength calibrated using 
ArXe arc lines. We finally combined the spectra at both AB positions and 
corrected for telluric absorption lines by dividing them by the A3V star 
HR8840, observed at a similar airmass, and multiplying by a blackbody of the 
corresponding effective temperature of 8500\,K. Spectroscopic data of 
LP 412-31 were reduced and analized in a similar way to HD\,221356\,D, 
but telluric correction was done using the A3V star HR\,1036, also observed at 
a similar airmass. Spectra of HD\,221356\,D and LP 412-31 in comparison 
with other standard objects and the main spectroscopic features are shown in 
Figure~\ref{fig:3}.  

HD\,221356\,D displays stronger water vapor absorption bands than the
standard M8 dwarf observed with the same instrumental setup and sky
conditions and reduced in the same manner as our target. This implies that
HD\,221356\,D has a cooler spectral type, quite likely within the L domain.
Aimed at deriving the spectral type of HD 221356D, we have
compared its LIRIS spectrum with data extracted from the IRTF libraries
\citep{2005ApJ...623.1115C}, see Fig. 3. 
We note that these data correspond mostly to dwarfs with solar composition.
The overall $HK$ spectral energy distribution (SED) of HD\,221356D is better 
reproduced by a spectral type of L0-L1. However we determine 
a spectral type of L1--L3 if we consider the different water indexes at 
$\sim$ 1.7 or 2.0 ~$\mu$m \citep{2001ApJ...552L.147T, 2007ApJ...657..511A, 2004ApJ...610.1045S}. 
The differences can be explained if the object is slightly metal poor (consistent with the primary), 
because the $K$-band flux should be reduced by the $H_2$ collisional induced absorption.
The strength of the sodium feature at 2.2 microns in object D compares better with solar metallicity
M9 dwarf than with the early L-type templates, however we prefer to rely on the general SED, 
which is better fit by L1 templates, than on a single feature that can be uncertain bearing in mind the 
poorly understood effects of low metallicity.

We also note that the water vapour absorption band at $\sim$ 1.5 ~$\mu$\,m, in the blue part of the 
$H$-band is more intense in this object than in early L dwarfs and more resembles mid/late L dwarfs. 
This is not an instrumental effect since it does 
not appear in the M8.5 object, but it may be due to a larger contamination by the primary at blue 
wavelengths than at red wavelengths. We therefore cannot, with confidence, 
assign this feature as unusual for this object. In summary, we  
estimate that HD\,221356\,D  is a slightly blue early L dwarf with a spectral type of L0--L2.  

\begin{figure*}
\centering
\includegraphics[scale=0.8]{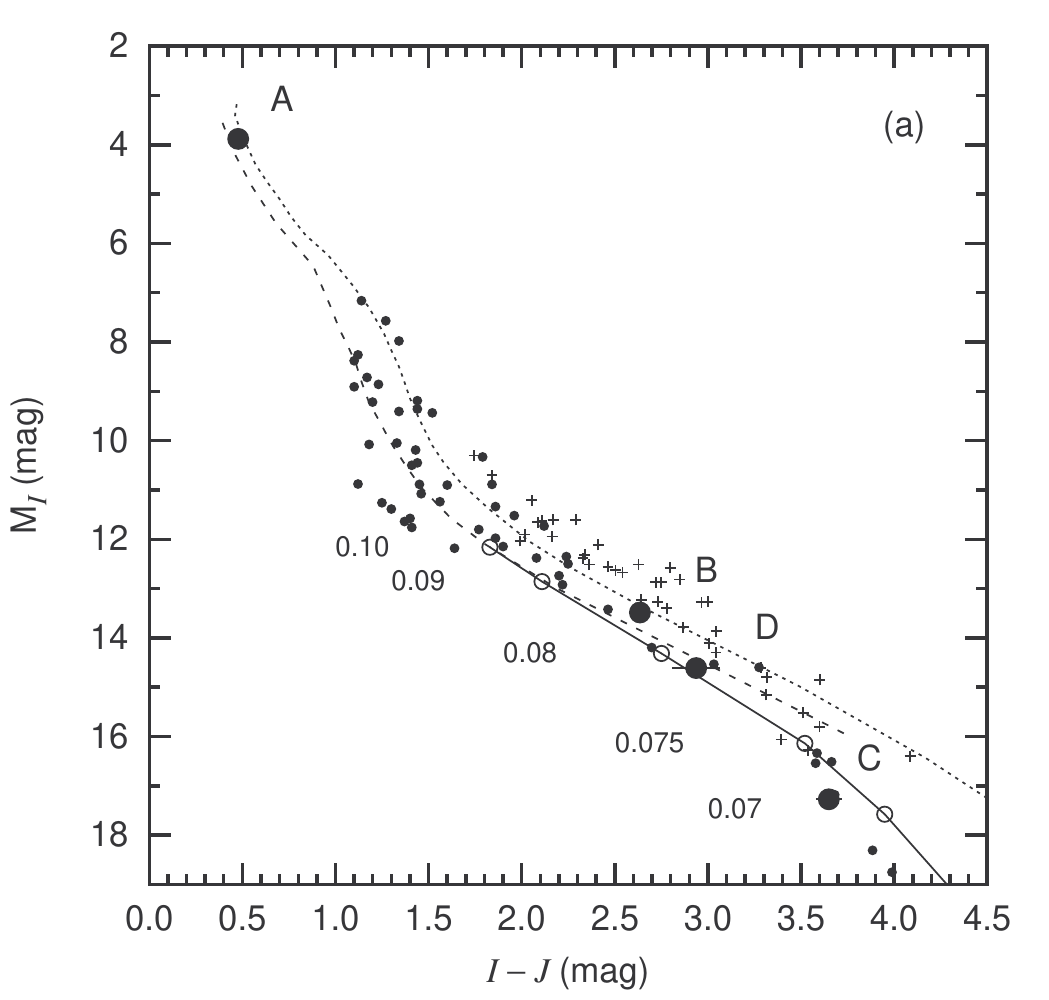}
\includegraphics[scale=0.825, trim=0pt 10pt 0pt 0pt]{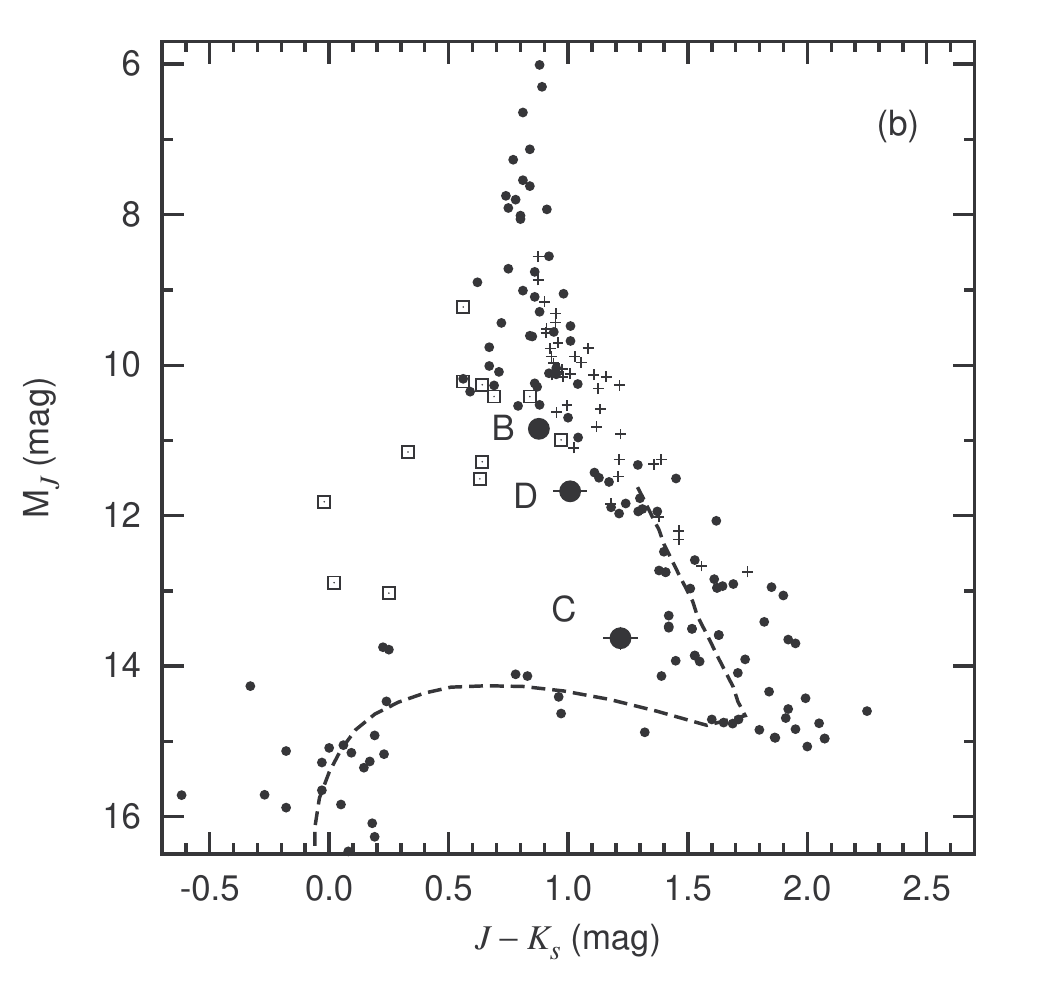}
\caption{Left panel: $M_I$, $I-J$ colour-magnitude diagram of the HD\,221356 system. 
The positions of the four components are marked with points and labelled with 
the corresponding letters. Pleiades low-mass stars and brown dwarfs (crosses) from \protect\cite{2010AA...519A..93B},
M and L dwarfs from \protect\cite{2006PASP..118..659L} with available parallaxes 
and M dwarfs (dots) from \protect\cite{2000ApJ...535..965L} are also plotted. 
The 5 Gyr isochrones of the NextGen models for solar and 
low metallicity stars \citep{1998A&A...337..403B}, represented by a dotted and dashed line, 
respectively, and of the DUSTY model for solar metallicity \citep{2000ApJ...542..464C}, 
shown as a solid line, are also included. The masses in solar mass units from the DUSTY 
model are also indicated and marked with open circles in the corresponding isochrone. 
Right panel: $M_J$, $J-K$ of BCD components of the HD\,221356 system. We have also added
L and T dwarfs from \protect\cite{2004AJ....127.2948V} and \protect\cite{2011ApJS..197...19K} and 
12 subdwarfs with measured parallaxes from \protect\cite{2012ApJ...752...56F}, depicted by squares. 
The mean L and T near-infrared photometric sequence from \protect\cite{2004AJ....127.2948V} 
is represented by a dashed line.}
\label{fig:4}
\end{figure*}
%
\section{Physical properties of HD 221356D}
%
Assuming a distance of 26.12$\pm$0.37 pc to the system, we calculated the absolute 
magnitudes of individual components and constructed the $\left(M_I,~I-J\right)$ 
and $\left(M_J,~J-K_s\right)$ colour-magnitude diagrams, shown in Fig. \ref{fig:4}. 
All four objects clearly follow a well defined photometric sequence, 
with the new companion located between the known B and C components. All the 
colours and magnitudes of the new object are in good agreement with its physical 
membership to the HD 221356 system. To better illustrate the position of the 
three low mass companions, we added in both panels of Fig. \ref{fig:4} the Pleiades 
low-mass stars and brown dwarfs (at $\rm{d=120~pc}$) from \cite{2010AA...519A..93B},
M and L dwarfs from \cite{2006PASP..118..659L} with available parallaxes and  
the field M dwarfs from \cite{2000ApJ...535..965L}. The Pleiades cluster offers a
homogeneous collection of objects with similar age and metallicity.
The right panel also includes L and T dwarfs from \cite{2004AJ....127.2948V}, 
\cite{2011ApJS..197...19K}, and 12 subdwarfs with spectral types
from M7 to L7 with measured parallaxes \citep{2012ApJ...752...56F}. 
In the $\left(M_I,~I-J\right)$ diagram we have also plotted the 5 Gyr isochrones 
of the NextGen models for solar and low metallicity $(\rm{[M/H]}=-0.5)$ 
\citep{1998A&A...337..403B} and of the DUSTY model for solar metallicity 
\citep{2000ApJ...542..464C}. 

We have depicted a veriety of objects with similar spectral types, that span from 
young ages ($\sim$120 Myr) and solar composition, to old stars with solar metallicities
and field subdwarfs with metal-poor abundances ([Fe/H]$\sim-0.5$, \citealp{2007ApJ...669.1235L}).
The low-mass components of the HD\,221356 system have slightly bluer colours than typical 
field stars, placing them on the blue edge of the photometric sequence defined 
by late M and L dwarfs.
However their colours are redder than known field subdwarfs, implying an intermediate
metallicity, being thus in good agreement with the metallicity determination of the primary.
 The $J-K_s$ and $I-J$ colours of the new companion, 
which are $1.01\pm0.06$ and $2.94\pm0.14$ mag respectively, correspond to the 
typical values of M7--M9 field dwarfs\citep{2002ApJ...564..452L, 1994AJ....107..333K}. 
These photometric colours suggest an earlier spectral type than that determined 
in the spectroscopic analysis. We attribute this difference to the sub-solar 
metallicity of the system, which affects the spectral energy distribution. 
In particular, the flux suppression in the $K_s$-band was already 
recognized in a group of L dwarfs, as a low metallicity feature 
\citep{2010ApJS..190..100K, 2011ASPC..448..531W}. Because of that, the spectral 
classification of the wide binary components of the HD\,221356 system, which is based on 
photometry, may be uncertain.
\begin{table}
\caption{Physical properties of low mass companions in HD 221356 system.}
\centering
\begin{tabular}{c c c c c c c}
\hline
\hline
Comp. & SpType & $\log~L/L_{\odot}$ &  $\rm{T_{eff}}$ (K) & $\rm{M/M_{\odot}}$ \\
\hline
 B & $\rm M8.0\pm1.5$ & $-3.28\pm0.09$ & 2351$-$2717 & $0.090\pm0.008$ \\
 C & $\rm L3.0\pm1.5$ & $-4.35\pm0.09$ & 1756$-$2141 & $0.065^{+0.007}_{-0.016}$ \\
 D & $\rm L1.0\pm1.0$ & $-3.61\pm0.10$ & 2079$-$2304 & $0.079\pm0.006$ \\
\hline
\end{tabular}
\label{tab:tab4}
\end{table}

The luminosities of the B, C and D components were derived from their 
$JHK_s$-band magnitudes, using the trigonometric distance of the primary,   
bolometric corrections from \cite{2004AJ....127.3516G} and spectral 
type-colour relations from \cite{2004AJ....127.2948V}. The effective 
temperature ranges were calculated adopting the temperature scale for 
high-gravity field dwarfs given by \cite{2004AJ....127.3516G}, assuming 
a spectral type range of L0--L2 for the new companion and M6.5--M9.5, 
L1.5--L4.5 for B and C, respectively. The resulting values are given in 
Table \ref{tab:tab4}. The masses of the new companion and B and C 
components were estimated using the DUSTY model from the Lyon group 
\citep{2000ApJ...542..464C}, which is only available for solar metallicity. 
We adopted a wide range of ages, using the 1, 5 and 10 Gyr isochrones. 
Masses were derived from their luminosities by interpolating the 
mass-luminosity relations given in the models.  
The differences in mass calculated for different ages are lower than the errors in mass 
determination resulting from the uncertainties of luminosities. For the 
new companion we finally adopted a mass of $0.079\pm0.006 M_{\odot}$, which 
is the average value obtained using $JHK_s$-band magnitudes, 
a spectral type range of L0$-$L2, and an age range of 1$-$10 Gyr. To take account 
of the differences in models for low metallicity stars, we have checked the mass 
of object D using the 5\,Gyr NextGen model for $[M/H]=-0.5$ \citep{1998A&A...337..403B}. We obtained
a mass of $0.083\pm0.002$, which is slightly larger, but still 
within the uncertainties of that determined for solar metallicity models.
%
\section{A benchmark sub-solar metallicity multiple system}
%
The age of the primary star in the HD\,221356 multiple system was estimated by 
\cite{2005ApJS..159..141V} to be 2.5$-$7.9 
Gyrs based on isochrone analysis. The lithium abundance of the primary star 
($\log n(Li)=2.5$ in the usual scale of $\log n(H)=12$) is typical of late F-type 
stars in clusters with ages in the 2$-$8 Gyr range \citep{2005A&A...442..615S}. 
The chromospheric activity of the star is also typical 
of a moderately old main sequence star \citep{2005ApJS..159..141V}.

HD 221356 is a slightly metal-poor stellar system whose components have masses just 
above and below the hydrogen burning limit. For sub-solar metallicities the 
stellar--brown dwarf borderline is expected to be shifted to higher masses, e.g. 
to $\sim0.079~M_{\odot}$ at $[M/H]=-0.5$ ($0.072~M_{\odot}$ at $[M/H]=0$) 
\citep{1998A&A...337..403B}.
In such a coeval, old system it becomes particularly interesting to investigate 
the lithium abundances of the very low mass components. While the M8 star (component B)
and the L0$-$L2 (component D) should have fully burnt their original lithium, component 
C with 0.065 solar mass may have preserved some amount of the initial lithium content. 
Theoretical models for solar metallicity predict full lithium depletion for such a mass,
however this may not be the case for sub-solar metallicity, since models also predict 
a less efficient depletion at low metallicities \citep{1997A&A...327.1039C}.
Observations of the Li abundance in the three low-mass components of this system will constrain
both the evolutionary models and the age of the system.

Multi epoch measurements of the system will allow to detect the orbital motion 
of companion D. Although the estimated orbital period (assuming 
circular orbit) is of the order of 5500 years, the relative change of the 
position would be up to $\sim 14~\rm mas/yr$, which is measurable using modern 
high spatial resolution imaging (e.g., adaptive optics, lucky imaging). The 
expected semi-amplitude of radial velocity variation of the primary induced 
by the presence of companion D will be of the order of 130 m/s (for 90$^\deg$ 
inclination), however the orbital period is too long to allow a full determination 
of the three-dimensional orbit. Maximum annual variations of roughly 0.15 m/s 
are expected, which may be explored with the new generation of ultra high 
precision spectrographs \citep{2012Natur.485..611W}. 
%
\section{Conclusions} 
%
Using the VHS and 2MASS surveys we have identified a new very low-mass 
companion (HD 221356 D) in the slightly metal-poor HD 221356 system, which thus becomes a 
quadruple. The new object is located 
at a projected distance of $\sim$ 312 AU from the F8 primary. 
The four components of the system follow a well defined photometric sequence.  
From near-infrared spectroscopy 
we determined L0--L2 spectral type for the D companion. Based on 
evolutionary models its mass is estimated at $0.079\pm0.006~\rm M_{\odot}$, 
and its effective temperature is in the range from 2100 to 2300 K. 
The $J-K_s$ and $I-J$ colours of the low-mass components are slightly bluer 
than field counterparts of the same spectral type. 
We interpret this as a result of the low metallicity of the system, which may 
become a reference for the spectral classification of metal poor M and L-type field objects. 
Since the distance and metallicity of the HD 221356 system are well known, 
the detailed study of its ultracool 
companions, which are located above and below the frontier 
between stars and brown dwarfs, can provide valuable constrains on evolutionary models and, 
in particular, shed light on the properties of objects on the transition from stellar to 
substellar regime. 

%
{\bf Acknowledgements.}
%
Based on observations obtained as part of the VISTA Hemisphere Survey, ESO Program; 179.A-2010 (PI: McMahon).
The VISTA Data Flow System pipeline processing and science archive are described in \cite{2004SPIE.5493..411I}
and \cite{2009MNRAS.399.1730C}.
This article is based on observations made with the TCS and IAC80 telescope operated on 
the island of Tenerife by the IAC in the Spanish Observatorio del Teide,
and with the William Herschel Telescope operated on the island of 
La Palma by the Isaac Newton Group in the Spanish Observatorio del Roque de los Muchachos 
of the Instituto de Astrof\'{i}sica de Canarias. This publication makes use of data 
products from the Two Micron All Sky Survey, which 
is a joint project of the University of Massachusetts and the Infrared Processing 
and Analysis Center/California Institute of Technology, funded by the National 
Aeronautics and Space Administration and the National Science Foundation. This 
research has benefitted from the SpeX Prism Spectral Libraries, maintained by 
Adam Burgasser at http://pono.ucsd.edu/~adam/browndwarfs/spexprism. 
This research has been supported by the Spanish Ministry
of Economics and Competitiveness under the projects
AYA2010-21308-C3-02, AYA2010-21308-C03-03 and AYA2010-20535. 
V.J.S.B. and N. L. are partially supported by the Spanish Ram\'{o}n y Cajal program.
%
\bibliographystyle{mn2e}
\bibliography{ms}

\begin{thebibliography}{}

\bibitem[\protect\citeauthoryear{{Allers} \& {et al.}}{{Allers} \& {et
  al.}}{2007}]{2007ApJ...657..511A}
{Allers} K.~N.,  {et al.} 2007, ApJ, 657, 511

\bibitem[\protect\citeauthoryear{{Baraffe}, {Chabrier}, {Allard} \&
  {Hauschildt}}{{Baraffe} et~al.}{1998}]{1998A&A...337..403B}
{Baraffe} I.,  {Chabrier} G.,  {Allard} F.,    {Hauschildt} P.~H.,  1998, A\&A,
  337, 403

\bibitem[\protect\citeauthoryear{{Bean}, {Sneden}, {Hauschildt}, {Johns-Krull}
  \& {Benedict}}{{Bean} et~al.}{2006}]{2006ApJ...652.1604B}
{Bean} J.~L.,  {Sneden} C.,  {Hauschildt} P.~H.,  {Johns-Krull} C.~M.,
  {Benedict} G.~F.,  2006, ApJ, 652, 1604

\bibitem[\protect\citeauthoryear{{Bihain}, {Rebolo}, {Zapatero Osorio},
  {B{\'e}jar} \& {Caballero}}{{Bihain} et~al.}{2010}]{2010AA...519A..93B}
{Bihain} G.,  {Rebolo} R.,  {Zapatero Osorio} M.~R.,  {B{\'e}jar} V.~J.~S.,
  {Caballero} J.~A.,  2010, A\&A, 519, A93

\bibitem[\protect\citeauthoryear{{Bonfils}, {Delfosse}, {Udry}, {Santos},
  {Forveille} \& {S{\'e}gransan}}{{Bonfils} et~al.}{2005}]{2005A&A...442..635B}
{Bonfils} X.,  {Delfosse} X.,  {Udry} S.,  {Santos} N.~C.,  {Forveille} T.,
  {S{\'e}gransan} D.,  2005, A\&A, 442, 635

\bibitem[\protect\citeauthoryear{{Burgasser}, {Reid}, {Siegler}, {Close},
  {Allen}, {Lowrance} \& {Gizis}}{{Burgasser}
  et~al.}{2007}]{2007prpl.conf..427B}
{Burgasser} A.~J.,  {Reid} I.~N.,  {Siegler} N.,  {Close} L.,  {Allen} P.,
  {Lowrance} P.,    {Gizis} J.,  2007, Protostars and Planets V, pp 427--441

\bibitem[\protect\citeauthoryear{{Burrows},  \& {et al.}}{{Burrows}
  et~al.}{1997}]{1997ApJ...491..856B}
{Burrows} A.,     {et al.} 1997, ApJ, 491, 856

\bibitem[\protect\citeauthoryear{{Burrows}, {Hubbard}, {Lunine} \&
  {Liebert}}{{Burrows} et~al.}{2001}]{2001RvMP...73..719B}
{Burrows} A.,  {Hubbard} W.~B.,  {Lunine} J.~I.,    {Liebert} J.,  2001,
  Reviews of Modern Physics, 73, 719

\bibitem[\protect\citeauthoryear{{Caballero}}{{Caballero}}{2007}]{2007ApJ...66%
7..520C}
{Caballero} J.~A.,  2007, ApJ, 667, 520

\bibitem[\protect\citeauthoryear{{Chabrier} \& {Baraffe}}{{Chabrier} \&
  {Baraffe}}{1997}]{1997A&A...327.1039C}
{Chabrier} G.,  {Baraffe} I.,  1997, A\&A, 327, 1039

\bibitem[\protect\citeauthoryear{{Chabrier}, {Baraffe}, {Allard} \&
  {Hauschildt}}{{Chabrier} et~al.}{2000}]{2000ApJ...542..464C}
{Chabrier} G.,  {Baraffe} I.,  {Allard} F.,    {Hauschildt} P.,  2000, ApJ,
  542, 464

\bibitem[\protect\citeauthoryear{{Close}, {Siegler}, {Potter}, {Brandner} \&
  {Liebert}}{{Close} et~al.}{2002}]{2002ApJ...567L..53C}
{Close} L.~M.,  {Siegler} N.,  {Potter} D.,  {Brandner} W.,    {Liebert} J.,
  2002, ApJ, 567, L53

\bibitem[\protect\citeauthoryear{{Costado}, {B{\'e}jar}, {Caballero}, {Rebolo},
  {Acosta-Pulido} \& {Manchado}}{{Costado} et~al.}{2005}]{2005A&A...443.1021C}
{Costado} M.~T.,  {B{\'e}jar} V.~J.~S.,  {Caballero} J.~A.,  {Rebolo} R.,
  {Acosta-Pulido} J.,    {Manchado} A.,  2005, A\&A, 443, 1021

\bibitem[\protect\citeauthoryear{{Cross}, {Collins}, {Hambly}, {Blake}, {Read},
  {Sutorius}, {Mann} \& {Williams}}{{Cross} et~al.}{2009}]{2009MNRAS.399.1730C}
{Cross} N.~J.~G.,  {Collins} R.~S.,  {Hambly} N.~C.,  {Blake} R.~P.,  {Read}
  M.~A.,  {Sutorius} E.~T.~W.,  {Mann} R.~G.,    {Williams} P.~M.,  2009,
  MNRAS, 399, 1730

\bibitem[\protect\citeauthoryear{{Cushing}, {Rayner} \& {Vacca}}{{Cushing}
  et~al.}{2005}]{2005ApJ...623.1115C}
{Cushing} M.~C.,  {Rayner} J.~T.,    {Vacca} W.~D.,  2005, ApJ, 623, 1115

\bibitem[\protect\citeauthoryear{{Dupuy}, {Liu}, {Bowler}, {Cushing},
  {Helling}, {Witte} \& {Hauschildt}}{{Dupuy}
  et~al.}{2010}]{2010ApJ...721.1725D}
{Dupuy} T.~J.,  {Liu} M.~C.,  {Bowler} B.~P.,  {Cushing} M.~C.,  {Helling} C.,
  {Witte} S.,    {Hauschildt} P.,  2010, ApJ, 721, 1725

\bibitem[\protect\citeauthoryear{{Emerson}, {McPherson} \&
  {Sutherland}}{{Emerson} et~al.}{2006}]{2006Msngr.126...41E}
{Emerson} J.,  {McPherson} A.,    {Sutherland} W.,  2006, The Messenger, 126,
  41

\bibitem[\protect\citeauthoryear{{Epchtein} \& {et al.}}{{Epchtein} \& {et
  al.}}{1999}]{1999A&A...349..236E}
{Epchtein} N.,  {et al.} 1999, A\%A, 349, 236

\bibitem[\protect\citeauthoryear{{Faherty}, {Burgasser}, {West}, {Bochanski},
  {Cruz}, {Shara} \& {Walter}}{{Faherty} et~al.}{2010}]{2010AJ....139..176F}
{Faherty} J.~K.,  {Burgasser} A.~J.,  {West} A.~A.,  {Bochanski} J.~J.,  {Cruz}
  K.~L.,  {Shara} M.~M.,    {Walter} F.~M.,  2010, AJ, 139, 176

\bibitem[\protect\citeauthoryear{{Faherty} \& {et al.}}{{Faherty} \& {et
  al.}}{2012}]{2012ApJ...752...56F}
{Faherty} J.~K.,  {et al.} 2012, ApJ, 752, 56

\bibitem[\protect\citeauthoryear{{Gizis}, {Monet}, {Reid}, {Kirkpatrick},
  {Liebert} \& {Williams}}{{Gizis} et~al.}{2000}]{2000AJ....120.1085G}
{Gizis} J.~E.,  {Monet} D.~G.,  {Reid} I.~N.,  {Kirkpatrick} J.~D.,  {Liebert}
  J.,    {Williams} R.~J.,  2000, AJ, 120, 1085

\bibitem[\protect\citeauthoryear{{Gizis}, {Reid}, {Knapp}, {Liebert},
  {Kirkpatrick}, {Koerner} \& {Burgasser}}{{Gizis}
  et~al.}{2003}]{2003AJ....125.3302G}
{Gizis} J.~E.,  {Reid} I.~N.,  {Knapp} G.~R.,  {Liebert} J.,  {Kirkpatrick}
  J.~D.,  {Koerner} D.~W.,    {Burgasser} A.~J.,  2003, AJ, 125, 3302

\bibitem[\protect\citeauthoryear{{Golimowski} \& {et al.}}{{Golimowski} \& {et
  al.}}{2004}]{2004AJ....127.3516G}
{Golimowski} D.~A.,  {et al.} 2004, AJ, 127, 3516

\bibitem[\protect\citeauthoryear{{Irwin} \& {et al.}}{{Irwin} \& {et
  al.}}{2004}]{2004SPIE.5493..411I}
{Irwin} M.~J.,  {et al.} 2004, in {Quinn} P.~J.,  {Bridger} A.,  eds, Society
  of Photo-Optical Instrumentation Engineers (SPIE) Conference Series Vol.~5493
  of Society of Photo-Optical Instrumentation Engineers (SPIE) Conference
  Series, {VISTA data flow system: pipeline processing for WFCAM and VISTA}.
pp 411--422

\bibitem[\protect\citeauthoryear{{Kirkpatrick}, {Cushing} \&
  {Gelino}}{{Kirkpatrick} et~al.}{2011}]{2011ApJS..197...19K}
{Kirkpatrick} J.~D.,  {Cushing} M.~C.,    {Gelino} C.~R.,  2011, ApJS, 197, 19

\bibitem[\protect\citeauthoryear{{Kirkpatrick} \& {et al.}}{{Kirkpatrick} \&
  {et al.}}{2010}]{2010ApJS..190..100K}
{Kirkpatrick} J.~D.,  {et al.} 2010, ApJS, 190, 100

\bibitem[\protect\citeauthoryear{{Kirkpatrick}, {Henry} \&
  {Simons}}{{Kirkpatrick} et~al.}{1995}]{1995AJ....109..797K}
{Kirkpatrick} J.~D.,  {Henry} T.~J.,    {Simons} D.~A.,  1995, AJ, 109, 797

\bibitem[\protect\citeauthoryear{{Kirkpatrick} \& {McCarthy} Jr.}{{Kirkpatrick}
  \& {McCarthy}}{1994}]{1994AJ....107..333K}
{Kirkpatrick} J.~D.,  {McCarthy} Jr. D.~W.,  1994, AJ, 107, 333

\bibitem[\protect\citeauthoryear{{Kraus} \& {Hillenbrand}}{{Kraus} \&
  {Hillenbrand}}{2007}]{2007ApJ...662..413K}
{Kraus} A.~L.,  {Hillenbrand} L.~A.,  2007, ApJ, 662, 413

\bibitem[\protect\citeauthoryear{{Lafreni{\`e}re} \& {et al.}}{{Lafreni{\`e}re}
  \& {et al.}}{2007}]{2007ApJ...670.1367L}
{Lafreni{\`e}re} D.,  {et al.} 2007, ApJ, 670, 1367

\bibitem[\protect\citeauthoryear{{Landolt}}{{Landolt}}{1992}]{1992AJ....104..3%
40L}
{Landolt} A.~U.,  1992, AJ, 104, 340

\bibitem[\protect\citeauthoryear{{Leggett}, {Allard}, {Dahn}, {Hauschildt},
  {Kerr} \& {Rayner}}{{Leggett} et~al.}{2000}]{2000ApJ...535..965L}
{Leggett} S.~K.,  {Allard} F.,  {Dahn} C.,  {Hauschildt} P.~H.,  {Kerr} T.~H.,
    {Rayner} J.,  2000, ApJ, 535, 965

\bibitem[\protect\citeauthoryear{{Leggett} \& {et al.}}{{Leggett} \& {et
  al.}}{2002}]{2002ApJ...564..452L}
{Leggett} S.~K.,  {et al.} 2002, ApJ, 564, 452

\bibitem[\protect\citeauthoryear{{L{\'e}pine}, {Rich} \& {Shara}}{{L{\'e}pine}
  et~al.}{2007}]{2007ApJ...669.1235L}
{L{\'e}pine} S.,  {Rich} R.~M.,    {Shara} M.~M.,  2007, ApJ, 669, 1235

\bibitem[\protect\citeauthoryear{{Lewis}, {Irwin} \& {Bunclark}}{{Lewis}
  et~al.}{2010}]{2010ASPC..434...91L}
{Lewis} J.~R.,  {Irwin} M.,    {Bunclark} P.,  2010, in {Y.~Mizumoto,
  K.-I.~Morita, \& M.~Ohishi} ed., Astronomical Data Analysis Software and
  Systems XIX Vol.~434 of Astronomical Society of the Pacific Conference
  Series, {Pipeline Processing for VISTA}.
p.~91

\bibitem[\protect\citeauthoryear{{Liebert} \& {Gizis}}{{Liebert} \&
  {Gizis}}{2006}]{2006PASP..118..659L}
{Liebert} J.,  {Gizis} J.~E.,  2006, PASP, 118, 659

\bibitem[\protect\citeauthoryear{{McCarthy} \& {Zuckerman}}{{McCarthy} \&
  {Zuckerman}}{2004}]{2004AJ....127.2871M}
{McCarthy} C.,  {Zuckerman} B.,  2004, AJ, 127, 2871

\bibitem[\protect\citeauthoryear{{Oscoz} \& {et al.}}{{Oscoz} \& {et
  al.}}{2008}]{2008SPIE.7014E.137O}
{Oscoz} A.,  {et al.} 2008, in Society of Photo-Optical Instrumentation
  Engineers (SPIE) Conference Series Vol.~7014 of Society of Photo-Optical
  Instrumentation Engineers (SPIE) Conference Series, {FastCam: a new lucky
  imaging instrument for medium-sized telescopes}

\bibitem[\protect\citeauthoryear{{Pinfield} \& {et al.}}{{Pinfield} \& {et
  al.}}{2012}]{2012MNRAS.422.1922P}
{Pinfield} D.~J.,  {et al.} 2012, MNRAS, 422, 1922

\bibitem[\protect\citeauthoryear{{Pinfield}, {Jones}, {Lucas}, {Kendall},
  {Folkes}, {Day-Jones}, {Chappelle} \& {Steele}}{{Pinfield}
  et~al.}{2006}]{2006MNRAS.368.1281P}
{Pinfield} D.~J.,  {Jones} H.~R.~A.,  {Lucas} P.~W.,  {Kendall} T.~R.,
  {Folkes} S.~L.,  {Day-Jones} A.~C.,  {Chappelle} R.~J.,    {Steele} I.~A.,
  2006, MNRAS, 368, 1281

\bibitem[\protect\citeauthoryear{{Sestito} \& {Randich}}{{Sestito} \&
  {Randich}}{2005}]{2005A&A...442..615S}
{Sestito} P.,  {Randich} S.,  2005, A\&A, 442, 615

\bibitem[\protect\citeauthoryear{{Skrutskie} \& {et al.}}{{Skrutskie} \& {et
  al.}}{2006}]{2006AJ....131.1163S}
{Skrutskie} M.~F.,  {et al.} 2006, AJ, 131, 1163

\bibitem[\protect\citeauthoryear{{Slesnick}, {Hillenbrand} \&
  {Carpenter}}{{Slesnick} et~al.}{2004}]{2004ApJ...610.1045S}
{Slesnick} C.~L.,  {Hillenbrand} L.~A.,    {Carpenter} J.~M.,  2004, ApJ, 610,
  1045

\bibitem[\protect\citeauthoryear{{Testi} \& {et al.}}{{Testi} \& {et
  al.}}{2001}]{2001ApJ...552L.147T}
{Testi} L.,  {et al.} 2001, ApJL, 552, L147

\bibitem[\protect\citeauthoryear{{Valenti} \& {Fischer}}{{Valenti} \&
  {Fischer}}{2005}]{2005ApJS..159..141V}
{Valenti} J.~A.,  {Fischer} D.~A.,  2005, ApJS, 159, 141

\bibitem[\protect\citeauthoryear{{van Leeuwen}}{{van
  Leeuwen}}{2007}]{2007AA...474..653V}
{van Leeuwen} F.,  2007, A\&A, 474, 653

\bibitem[\protect\citeauthoryear{{Vrba} \& {et al.}}{{Vrba} \& {et
  al.}}{2004}]{2004AJ....127.2948V}
{Vrba} F.~J.,  {et al.} 2004, AJ, 127, 2948

\bibitem[\protect\citeauthoryear{{West}, {Bochanski}, {Bowler}, {Dotter},
  {Johnson}, {L{\'e}pine}, {Rojas-Ayala} \& {Schweitzer}}{{West}
  et~al.}{2011}]{2011ASPC..448..531W}
{West} A.~A.,  {Bochanski} J.~J.,  {Bowler} B.~P.,  {Dotter} A.,  {Johnson}
  J.~A.,  {L{\'e}pine} S.,  {Rojas-Ayala} B.,    {Schweitzer} A.,  2011, in
  {Johns-Krull} C.,  {Browning} M.~K.,   {West} A.~A.,  eds, 16th Cambridge
  Workshop on Cool Stars, Stellar Systems, and the Sun Vol.~448 of Astronomical
  Society of the Pacific Conference Series.
p.~531

\bibitem[\protect\citeauthoryear{{Wilken} \& {et al.}}{{Wilken} \& {et
  al.}}{2012}]{2012Natur.485..611W}
{Wilken} T.,  {et al.} 2012, Nature, 485, 611

\bibitem[\protect\citeauthoryear{{Zhang} \& {et al.}}{{Zhang} \& {et
  al.}}{2010}]{2010MNRAS.404.1817Z}
{Zhang} Z.~H.,  {et al.} 2010, MNRAS, 404, 1817

\end{thebibliography}
\end{document}